\documentclass[12pt]{article}
\usepackage{cite}
\usepackage{amsmath,latexsym,epsfig}
\usepackage{amsfonts,amssymb,amscd}

\newcommand{\grgen}{{\xi}}

\usepackage{amsmath}
\usepackage{amssymb}
\usepackage{epsfig}

\begin{document}

% macros
\def\cM{{\cal{M}}}
\def\cP{{\cal{P}}}
\def\cT{{\cal{T}}}
\def\cV{{\cal{V}}}
\def\cW{{\cal{W}}}

\def\bfi{{\rm\bf i}}
\def\bfu{{\rm\bf u}}
\def\bfx{{\rm\bf x}}
\def\bfy{{\rm\bf y}}
\def\bfM{{\rm\bf M}}

\def\bbP{{\mathbb{P}}}
\def\reals{{\mathbb{R}}}
\def\comp{{\mathbb{C}}}
\def\integ{{\mathbb{Z}}}
\def\tsh{{\textstyle{\frac{1}{2}}}}
\def\tsqu{{\textstyle{\frac{1}{4}}}}
\def\tr{{\rm{trace\,}}}
\def\rintn{\int_{\reals^N}\!\!}
\def\symgp{{\mathfrak{S}}}
\def\av#1{\left\langle#1\right\rangle_{\cW_N}}
\def\efM{e^{-\tsqu\tr\bfM^2}}
\def\etM{e^{-\tsh\tr\bfM^2}}
\def\ip#1#2{\left\langle #1,#2\right\rangle}
\def\avv#1{\left\langle#1\right\rangle_{\cV_N}}
\def\Fix{{\sf{Fix}}}
\def\ipt#1#2#3{\left\langle #1,#2\right\rangle_{#3}}
\def\stat{{\sf{stat}}}
\def\ones{{\bf 1}}
\def\rel{\,{\sf rel}\,\,}
\def\mybar{\overline}

\def\proof{{\rm\bf Proof:\quad}}
%-------------

%%%%%%%%%%%%%%%%%%%%%%%%%%%%%%%%%%%%%%%%%%%%%%%%%%%%%%%%%%%%%%%%%%%%%

\begin{titlepage}

% the footnote symbols are only redefined for the title page !
\renewcommand{\thefootnote}{\alph{footnote}}
\vspace*{-3.cm}
\begin{flushright}

\end{flushright}

\vspace*{0.3in}

{\begin{center} {\Large\bf  Generalized Penner  model and the Gaussian beta ensemble}

\end{center}}

\vspace*{.8cm} {\begin{center} {\large{\sc
               
                }}
\end{center}}
\vspace*{0cm} {\it
\begin{center}
 \vspace*{.8cm} {\begin{center} {\large{\sc
                Noureddine~Chair
                }}
\end{center}}
\vspace*{0cm} {\it
\begin{center}
 Physics Department,
The University of Jordan, Amman, Jordan 
 \begin{center}

     Email: n.chair@ju.edu.jo\\ \hspace{19mm}\\
\end{center}
\end{center}}
\end{center}}

\vspace*{1.5cm}

\begin{center} Abstract\end{center}

 \ \  In this paper,  a new expression  for the partition function of the generalized Penner  model given by  Goulden, Harer and Jackson is derived. The Penner and the orthogonal Penner partition functions are special cases of this formula. The parametrized Euler characteristic  $\xi^s_g(\gamma)$ deduced from our expression of the partition function is shown to exhibit a contribution from  the orbifold Euler characteristic of the moduli space of Riemann surfaces of genus $g$, with $s$ punctures,  for all parameters $\gamma$ and  $g$ odd. The other contributions for $g$ even are  linear combinations of the Bernoulli polynomials at rational arguments. It turns out that the free energy coefficients of the generalized Penner model in the  continuum limit, are identical to  those coefficients  in the large $N$ expansion of the  Gaussian $\beta$-ensemble. Moreover, the duality enjoyed by the generalized Penner model, is also  the  duality symmetry of the Gaussian $\beta$-ensemble. Finally, a shift  in the 't Hooft coupling constant required by  the refined topological string, would leave the  Gaussian $\beta$-ensemble duality  intact. This duality is identified with   the remarkable duality of the $c=1$ string at radius $R=\beta$.

\vspace*{.5cm}
\end{titlepage}

\renewcommand{\thefootnote}{\arabic{footnote}}

 \section{Intoduction}
 \ \  In their interesting paper, Goulden, Harer and Jackson \cite{ Harer1} generalized the Penner matrix model \cite{ Penner},  and obtained an expression for the parametrized Euler characteristic  $\xi^s_g(\gamma)$.  This polynomial in $\gamma^{-1}$ gives  when specializing the parameter, $\gamma$, to  $\gamma=1$ and  $\gamma=1/2$,  the orbifold Euler characteristic of the moduli space of complex algebraic curves ( Riemann surfaces ) of genus $g$  with $s$ punctures and  real algebraic curves (non-orientable surfaces) of genus  $g$  with $s$ punctures, respectively. It was shown explicitly that for odd $g$, $\xi^s_g(1/2)$   coincides with  the orbifold Euler characteristic of the moduli space of complex algebraic curves \cite{Harer2}, \cite{Penner}.  On the other hand, if $g$ is even,  $\xi^s_g(1/2)$ corresponds  to the orbifold Euler characteristic of the moduli space of real algebraic curves,  also known as  the orthogonal Penner model  \cite{Chekhov}, \cite{Chair1}. One must say that the Penner approach  is more accessible to physicists since it uses  Feynman diagrams and random matrices \cite{Bessis}.
 
 \ \  In this paper we give  an alternative  formula for the partition function from which $\xi^s_g (\gamma)$ can be computed. The simplicity of this formula is that  the partition functions of the Penner and the orthogonal Penner models are transparent, and so  this formula may be considered as a parametrized partition function for the generalized Penner model. This formula  shows that for odd $g$, the  parametrized Euler characteristic  $\xi^s_g(\gamma)$ is a sum of two terms,  the orbifold Euler characteristic of the moduli space of complex algebraic curves and a  linear combination   of the Bernoulli polynomials at rational arguments. For even $g$,  $\xi^s_g(\gamma)$ is shown to  coincide with the results obtained  previously  in \cite{Harer1}.
 
 \ \  We find that in the continuum limit (double scaling), both the Penner and the generalized Penner  matrix models have the same critical points. The free energy of the latter  model in this limit is related to the  $c=1$ string free energy at radius $R=\gamma$ \cite{Gross}. This observation was made recently in connection with  $\mathcal{N} = 2$  gauge theory in the $\Omega$-background \cite{Walcher}. It is also interesting to note that the free energy in the continuum limit, and the  nonperturbative terms \footnote{Here, these terms arise from the volume of the gauge group only, that is,  instanton corrections are turned-off. If, however,   these corrections are taken into account, then the matrix model partition function would  be  defined in the complex plane  with a suitable contour of integration \cite{Marino2}.} of the present model are given by  the free energy of the Gaussian $\beta$-ensemble in the large $N$ limit. This follows from the expression for the partition function of the generalized Penner matrix model, see Eq. (\ref{p0}), below. This can be understood  from the fact that  matrix models are considered as topological  gauge theories, like the Chern-Simons gauge theory. In particular, the  nonperturbative terms of the $SU(N)$-Chern-Simons gauge theory, that is, the volume of the  $SU(N)$ gauge group \cite{Ooguri}, may be reproduced by taking the double scaling limit of the perturbative $SU(N)$-Chern-Simons gauge theory \cite{Das}. Note that the double scaled theory  also corresponds to the $c=1$ matrix model at self-dual radius, that is, the Penner model or equivalently, closed topological (B model) strings on $ S ^{3}$ deformation of the conifold \cite{Ghoshal}. The nonperturbative terms in the Penner model may be shown to be captured by the continuum limit of the Penner model itself. This was extended  to $SO/Sp$-Chern-Simons Gauge Theories in  \cite{Chair2}.
 
 \ \  These aforementioned observations show clearly that the Gaussian $\beta$-ensemble partition  function may be considered as a volume of a certain gauge group of the generalized Penner matrix model. For example, the partition  function for $\beta=1$, gives the volume of  the $SU(N)$ gauge group, while for $\beta=1/2, 2 $ the corresponding gauge groups are the $SO(N), Sp(N)$ respectively.  Therefore the generalized Penner matrix model in the  continuum limit may be considered  as an  alternative approach  to carry out the large $N$-expansion  of the Gaussian $\beta$-ensemble  free energy \cite {Spreafico,Walcher,Brini,Krefl}. In particular,  the free energy in the continuum limit of this model is the asymptotic expansion of the
Barnes double-Gamma function \cite{Spreafico}.

\ \ The  explicit expression for the free energy of the generalized Penner model in the continuum limit ( see Eq.(37))  bears the  duality symmetry $ \gamma \rightarrow1/\gamma$ and $\mu \rightarrow -\gamma \mu$, where  $\mu=N(1-t)$ is  a continuum parameter. This duality symmetry is  similar to the one  found earlier in \cite{Mulase},  which may be considered as a natural generalization of the equivalence of $Sp(2N)$ and $SO(−2N)$ gauge theories \cite{Mkrtchyan}.
As a result, the Gauassian $\beta$-ensemble free energy in the large N limit should be invariant
under the duality transformation $\beta \rightarrow 1/\beta$ and $t \rightarrow -\beta t$, where $t$ is  the 't Hooft
coupling. For $\beta=1$, one recovers the $c=1$ string at self-radius, i.e., the topological $B$-model string on the deformed conifold. This is known to be even in the string  coupling constant $g_{s}$.  Shifting  the coupling  constant $t$ required by the refined topological string \cite{Iqbal,Aganagic},  the duality symmetry is kept intact,  and coincides with  the  remarkable duality symmetry of the $c=1$ string at radius $R=\beta$ \cite{Gross}.

 \ \ In the following sections, we proceed as follows: in section 2, we   derive   formulas for the partition function of the generalized Penner matrix model, and the parametrized Euler characteristic  $\xi^s_g(\gamma)$. A  brief review of the  Gaussian $\beta$-ensemble  and  its connection to the generalized Penner model is given in section 3. Here, the double-scaling  limit is carried out and is shown to have the same critical points as in the usual Penner model. Also, it is shown that the generalized Penner model in the continuum limit reproduces the large $N$-expansion  of the Gaussian $\beta$-ensemble  free energy. The duality symmetry of the generalized Penner model is discussed  in section 4, and is shown  to induce the  same duality symmetry in  the Gaussian $\beta$-ensemble, where it  holds order by order. Finally, the conclusion of this work is drawn in section 5.

 \section{The generalized Penner partition function and the parametrized Euler characteristic }
 
 \ \  The partition function for the generalized Penner  model \cite{Harer1}  can be
  written as
 \begin{eqnarray}
\label{p0} 
W_\gamma(N,t)=
\frac{
\displaystyle{
\int_{\reals^N} \vert \Delta(\lambda)\vert^{2\gamma}
\prod_{j=1}^N e^{-i\gamma\lambda_j/ \sqrt{t}}
e^{-\frac{\gamma}{t}\log(1-i\sqrt{t}\lambda_j)}d\lambda_{j}
}
}{
\displaystyle{
\int_{\reals^N} \vert \Delta(\lambda)\vert^{2\gamma}\prod_{j=1}^Ne^{-\gamma \sum_{i=1}^{N}\lambda_{i}^2/2}d\lambda_{j}}},
\end{eqnarray}
where $\Delta(\lambda)= \prod_{1\le i<j\le N}(\lambda_j-\lambda_i)$ is  the Vandermonde determinant. If we set $\gamma=1$, then $W_1(N,t)$ is the Penner model partition function \cite{Penner}, and so this generalized model may be considered  as a deformed Penner model; the deformation parameter being $\gamma$. In this model, the partition function  of the Gaussian $\beta$-ensemble  that appears in the denominator  plays the role of the volume of certain gauge group. This corresponds to the nonperturbative terms in the generalized Penner model.

\ \  In the next section, we will check that the free energy for this model in the continuum limit  is the large $N$ expansion for the free energy of the Gaussian $\beta$-ensemble.  
 The parametrized Euler characteristic  \cite{Harer1} was shown to be  connected to the partition function $W_\gamma(N,t)$ through the following expression 
\begin{eqnarray}
\label{p1}
{\grgen}^s_g (\gamma)= s!(-1)^s[N^s t^{g+s-1}]\,
\frac{1}{\gamma}\log W_\gamma(N,t),
\end{eqnarray}
where $ [X]Y$  is a short notation for the coefficient of $ X$ in the expansion of $Y $, and  $$W_\gamma(N,t) =\left(\sqrt{2\pi}\,\, e^{-\gamma/t}
\left(\frac{\gamma}{t}\right)^{\frac{\gamma}{t}-\tsh((N-1)\gamma+1)}\right)^N
\prod_{j=0}^{N-1}\frac{1}{\Gamma\left(\frac{\gamma}{t}-\gamma j\right)},$$
is the partition function  obtained from Eq. (\ref{p0}) using the Selberg integration formula.

\ \ 
 To motivate our method in obtaining  $\xi^s_g(\gamma)$,  let us  consider the case in which $\gamma=1/2$. To that end,  we  use the Legendre duplication formula to  show that 
\begin{equation}
\label{p2}
\prod_{j=0}^{N-1}\Gamma\left(\frac{1}{2t}-\frac{j}{2}\right)=\prod_{j=0}^{N/2-1}\frac{\sqrt{\pi}}{2^{1/t-(N-(2j+1))-1}}\Gamma\left(\frac{1}{t}-(N-(2j+1))\right),
\end{equation}
and from the identity
\begin{equation}
\label{p3}
\Gamma\left(\frac{1}{t}-(N-(2j+1))\right)=\frac{t^{N-(2j+1)}\Gamma\left(\frac{1}{t}\right)}{\prod_{p=1}^{N-(2j+1)}(1-pt)},
\end{equation}
we get
 \begin{eqnarray}
 \label{p4}
 W_{1/2}(N,t) =\left(\frac{\sqrt{2\pi t}\,\, (et)^{-1/t}}{\Gamma\left(\frac{1}{t}\right)}
\right)^{N/2}\prod_{j=0}^{N/2-1}\prod_{p=1}^{N-(2j+1)}(1-pt).
 \end{eqnarray}
By setting $\gamma=1/q$,  $q$ being an integer and $N=qK$, Goulden, Harer and Jackson have derived the following formula
\begin{eqnarray}
\label{p5}
W_\frac{1}{q}(qK,t)=
\left(\frac{\sqrt{2\pi t}
}{ \Gamma(\frac{1}{t})\,(et)^\frac{1}{t}
}\right)^K
\frac{{\prod_{l=1}^K\prod_{j=1}^{ql}(1-jt)}
}{ {\prod_{j=1}^K(1-tqj)}},
\end{eqnarray}
and in particular, 
 \begin{eqnarray}
 \label{p6}
 W_{1/2}(N,t) =\left(\frac{\sqrt{2\pi t}\,\, (et)^{-1/t}}{\Gamma\left(\frac{1}{t}\right)}
\right)^{N/2}\frac{{\prod_{l=1}^{N/2}\prod_{j=1}^{2l}(1-jt)}
}{ {\prod_{j=1}^{N/2}(1-2tj)}
}.
 \end{eqnarray}
 As a consequence, we deduce the following  identity 
\begin{eqnarray}
\label{p7}
\frac{{\prod_{l=1}^{N/2}\prod_{j=1}^{2l}(1-jt)}
}{ {\prod_{j=1}^{N/2}(1-2tj)}
}&=&\prod_{j=0}^{N/2-1}\prod_{p=1}^{N-(2j+1)}(1-pt)\nonumber\\&=&\prod_{p=1}^{N/2}(1-(2p-1)t)^{N/2-p+1}(1-(2p)t)^{N/2-p},
 \end{eqnarray}
 where $N$ is assumed to be even. Note that the last equality follows from the identity $$ \prod_{p=1}^{N-(2j+1)}(1-pt)=\prod_{p=1}^{N/2-j}(1-(2p-1)t)\prod_{p=1}^{N/2-1-j}(1-2pt).$$  As a result, the partition  function $ W_{1/2}(N,t) $ is reduced to a single product as follows
 \begin{eqnarray}
 \label{partition}
 W_{1/2}(N,t) =\left(\frac{\sqrt{2\pi t}\,\, (et)^{-1/t}}{\Gamma\left(\frac{1}{t}\right)}
\right)^{N/2}\prod_{p=1}^{N/2}(1-(2p-1)t)^{N/2-p+1}(1-(2p)t)^{N/2-p}.
 \end{eqnarray}
 Therefore, the free energy in this case reads
 \begin{eqnarray}
 \label{free energy}
2\log W_{1/2}(N,t) =\log\left(\left(\frac{\sqrt{2\pi t}\,\, (et)^{-1/t}}{\Gamma\left(\frac{1}{t}\right)}
\right)^{N}\prod_{p=1}^{N}(1-pt)^{N-p}\right)+\log\prod_{p=1}^{N/2}(1-(2p-1)t)
 \end{eqnarray}
 This is exactly  the free energy of  the  orthogonal Penner model \cite{Chair1}, where the first term represents the Penner free energy.
 
 \ \  If we  use the following  identity, the formula for the partition function  of the generalized Penner model for all  $q$'s, would involve  a single product  
\begin{eqnarray}
\label{p8}
\frac{{\prod_{l=1}^K\prod_{j=1}^{ql}(1-jt)}
}{ {\prod_{j=1}^K(1-tqj)}}&=&\prod_{j=0}^{N/q-1}\prod_{p=1}^{N-(qj+1)}(1-pt)\nonumber\\&=&\prod_{p=1}^{N/q}(1-(qp-(q-1))t)^{N/q-p+1}(1-(qp-(q-2))t)^{N/q-p+1} \nonumber\\&&(1-(qp-(q-3))t)^{N/q-p+1}\cdots (1-(qp)t)^{N/q-p}.
\end{eqnarray}
The products on the right-hand side are  taken over all non congruent and  congruent to $q$, of which   $q-1$ products are non congruent to $q$, and $N$ being a multiple of $q$. This identity which was first proposed and checked for several values of $N$, is a natural generalization of Eq. (\ref{p7}). The derivation of this identity is given in  appendix \ref{identity}. Using this  identity, one can write the partition function of the generalized Penner model in terms of a single product as
\begin{eqnarray}
\label{p9}
W_\frac{1}{q}(N,t)&&=\left(\frac{\sqrt{2\pi t}}{ \Gamma(\frac{1}{t})\,(et)^\frac{1}{t}}\right)^{N/q}\prod_{p=1}^{N/q}(1-(qp-(q-1))t)^{N/q-p+1}\nonumber\\&&(1-(qp-(q-2))t)^{N/q-p+1}(1-(qp-(q-3))t)^{N/q-p+1}\cdots\ (1-(qp)t)^{N/q-p}.\nonumber\\
\end{eqnarray}

\ \ Finally, the expression for the free energy reads 

\begin{eqnarray}
\label{p11}
q\log W_\frac{1}{q}(N,t)& =& \log\left(\frac{\sqrt{2\pi t}
}{ \Gamma(\frac{1}{t})\,(et)^\frac{1}{t}
}\right)^{N}+\sum_{p=1}^{N}\left(N-p\right)\log\left(1-pt\right)\nonumber\\&+&\sum_{p=1}^{N}\log\left(1-pt\right)-\sum_{p=1}^{N/q}\log\left(1-qpt\right)\nonumber\\&+&\sum_{p=1}^{N/q}\log\left(1-(qp-(q-2))t\right)\nonumber\\&+&2\sum_{p=1}^{N/q}\log\left(1-(qp-(q-3))t\right)+\cdots+(q-2)\sum_{p=1}^{N/q}\log\left(1-(qp-1)t\right).\nonumber\\  
\end{eqnarray} 
The first line  in this formula is nothing but the free energy of the Penner  model. This computes the orbifold Euler characteristic of the moduli space of  Riemann surfaces  of genus $g$  with $s$ punctures $\chi(\cM_g^s)$ \cite{Harer2}, \cite{Penner}. As a consequence,  the parametrized Euler characteristic $\xi^s_g(\gamma)$ for any $q\geq2$, contains a contribution coming from the orbifold Euler characteristic of the moduli space of complex algebraic curves
given by 
\begin{equation}
\label{orb Euler}
\chi(\cM_g^s) = (-1)^s\frac{(g+s-2)!}{(g+1)(g-1)!}B_{g+1},
\end{equation}
for  odd  $g$, and $B_{g}$ is the $g$th Bernoulli number.  One should note that the third and the fourth lines in Eq. (\ref{p11}) do contribute to  $\xi^s_g(\gamma)$ only for $q\geq3$.  The free energy in Eq. (\ref{p11})  gives the well known results  for $q=1$, $q=2$.
 
 \ \  Next, we  derive a suitable  expression  for the free energy $ q\log W_\frac{1}{q}(N,t)$ that computes  $\xi^s_g(\gamma)$,  such that the first line which generates  $\chi(\cM_g^s)$ is  omitted. For this purpose,  $ q\log W_\frac{1}{q}^{1}(N,t)$  is considered to be  the  free energy. Expanding the latter term, one has 
\begin{eqnarray}
\label{p12}
q\log W_\frac{1}{q}^{1}(N,t)&=&-\sum_{m\geq1}\frac{t^m}{m}\left(\sum_{p=1}^{N}p^{m}-\sum_{p=1}^{N/q}(qp)^{m}\right)-\sum_{m\geq1}\frac{t^m}{m} \sum_{j\geq0}^{m}(-1)^j\binom{m}{j}\sum_{p\geq1}^{N/q}q^{m}p^{m-j}\nonumber\\&\times&\left( (1-\frac{2}{q})^{j}+2(1-\frac{3}{q})^{j}+\cdots+(q-2)(\frac{1}{q})^{j}\right).
\end{eqnarray}
Using the power sum formula
$$\sum_{j=1}^{n}j^k=\frac{1}{k+1}\sum_{r=1}^{k+1}\binom{k+1}{r}B_{k+1-r}
(-1)^{k+1-r}n^r, $$ the free energy reads 
\begin{eqnarray}
\label{p13}
q\log W_\frac{1}{q}^{1}(N,t)&=&-\sum_{m\geq1}\frac{t^m}{m}\left(\frac{1}{m+1}\sum_{l=1}^{m+1}\binom{m+1}{l}(-1)^{m+1-l}B_{m+1-l}N^{l}\right)\nonumber\\&+&\sum_{m\geq1}\frac{t^m}{m}\left(\frac{q^m}{m+1}\sum_{l=1}^{m+1}\binom{m+1}{l}(-1)^{m+1-l}B_{m+1-l}q^{-l} N^{l}\right)\nonumber\\&-&\sum_{m\geq1}\frac{t^m}{m}\left(\sum_{j=0}^{m}(-1)^{j}\binom{m}{j}\frac{q^m}{m-j+1}\right)\nonumber\\&\times&\left(\sum_{l=1}^{m-j+1}\binom{m-j+1}{l}(-1)^{m-j+1-l}B_{m-j+1-l}q^{-l} N^{l}\right)\nonumber\\&\times&\left( (1-\frac{2}{q})^{j}+2(1-\frac{3}{q})^{j}+\cdots+(q-2)(\frac{1}{q})^{j}\right).
\end{eqnarray}

\ \ If we  extract  the coefficient $ \xi^s_g(\gamma) $ of $s!(-1)^{s}N^{s}t^{g+s-1} $ from Eq. (16), we get ( with $m=g+s-1$ and $l=s$) 
\begin{eqnarray}
\label{p14}
\xi^s_g(\gamma) &=& s!(-1)^{s}[N^{s}t^{g+s-1}]q\log W_\frac{1}{q}^{1}(N,t)=(-1)^{s+1}\frac{(g+s-2)!}{g!}(-1)^{g}\left(1-q^{g-1}\right)B_{g}\nonumber\\&+&(-1)^{s+1}\frac{(g+s-2)!}{g!}(-1)^{g}q^{g-1}\sum_{j=0}^{g}\binom{g}{j}B_{g-j}\nonumber\\&\times&\left( (1-\frac{2}{q})^{j}+2(1-\frac{3}{q})^{j}+\cdots+(q-2)(\frac{1}{q})^{j}\right).
\end{eqnarray}

\ \   For even  $g$, the explicit computation for $\xi^s_g(\gamma)$ which is carried out in Appendix \ref{PEc} leads to the following closed form formula
 \begin{eqnarray}
\label{p16}
 \xi^s_g(\gamma)&=&(-1)^{s+1}\frac{(g+s-2)!}{g!}\left(1-q^{g-1}\right)B_{g}\nonumber\\&+&(-1)^{s+1}\frac{(g+s-2)!}{g!}q^{g-1}\left( \frac{q-2}{2} \right)\left(\frac{1}{q^{g-1}}-1\right)B_{g}\nonumber\\&=&(-1)^{s}\frac{(g+s-2)!}{g!2}\left(q^g-q\right)B_{g}.
\end{eqnarray}
This result is indeed in complete agreement with the expression obtained  by Goulden, Harer, and Jackson \cite{Harer1}.

\ \  For odd $g$, the first line given in Eq. (\ref{p11}) generates the orbifold Euler characteristic for the moduli space of complex algebraic curves  of genus $g$  with $s$ punctures $\chi(\cM_g^s)$. The other contributions for  $\xi^s_g(\gamma)$ do  come only from the last term of Eq. (\ref{p14}) since the first term of latter equation does not contribute  for  odd $g$   and $g>1$. Hence,  
\begin{eqnarray}
\label{p17}
 \xi^s_g(\gamma)&=&(-1)^s\frac{(g+s-2)!}{(g+1)(g-1)!}B_{g+1}\nonumber\\&+&(-1)^{s}\frac{(g+s-2)!}{g!}q^{g-1}\left( B_{g}(1-\frac{2}{q})+2B_{g}(1-\frac{3}{q})+\cdots+(q-2)B_{g}(\frac{1}{q})\right),
 \end{eqnarray}
 where  $B_{n}(x)=\sum_{k=0}^{n}\binom{n}{k}B_{k}x^{n-k} $ is  the Bernoulli polynomial of degree $n$. 
 Making use of the symmetry $ B_{n}(1-x)=-B_{n}(x)$ for odd $n$ , the sum on the right-hand side  for odd $q$ can  be written as
 \begin{eqnarray}
\label{p18}
&&B_{g}(1-\frac{2}{q})+2B_{g}(1-\frac{3}{q})+\cdots+(q-2)B_{g}(\frac{1}{q})\nonumber\\&&=(q-2)B_{g}(\frac{1}{q})+(q-4)B_{g}(\frac{2}{q})+\cdots+B_{g}(\frac{q-1}{2q}). 
\end{eqnarray}
However,  if $q$ is even, Eq. (20) still holds except that the last term is replaced by $ 2B_{g}(\frac{q-2}{2q})$.

\ \  As a consequence, and for odd $g$ and $q$, the expression for the parametrized Euler characteristic becomes
\begin{eqnarray}
\label{p19}
 \xi^s_g(\gamma)&=&(-1)^s\frac{(2g+s-3)!}{(2g)(2g-2)!}B_{2g}\nonumber\\&+&(-1)^{s}\frac{(2g+s-3)!}{(2g-1)!}q^{2g-2}\left(\sum_{i=1}^{(q-1)/2}(q-2i)B_{2g-1}(\frac{i}{q})\right),\nonumber\\
 \end{eqnarray}
 
  while for even $q$, the maximum value of $i$ in the sum is $ (q-2)/2$. Therefore, the  parametrized Euler characteristic $ \xi^s_g(\gamma)$ exhibits a contribution from the orbifold Euler characteristic of the moduli space of complex algebraic curves $\chi(\cM_g^s)$. The other contributions  are linear combinations of the Bernoulli polynomials at rational arguments. Note that for the real algebraic curves case ($q=2$), the parametrized Euler  characteristic is equal to  $\chi(\cM_g^s)$ for odd $g$.
 
 \ \  Recall that the parametrized  Euler characteristic   $ {\grgen}^s_g (\gamma)$, is  coefficient of $  s!(-1)^sN^s t^{g+s-1}$ in the expansion of $\frac{1}{\gamma}\log W_\gamma(N,t)$. In terms of the generating series for the number of embedded graphs in a surface, where an edge and its  end are distinguished, these kind of graphs are called rooted  maps. The parametrized  generating series is given  by $$M_{\gamma}(\mathbf{y},x,z)= \sum_{\mathbf{I},j,n} m_{\gamma}({\mathbf {I}},x ^{j},z^{n})\mathbf{y}^{\mathbf {I}}x ^{j}z^{n},$$
 where  $ \mathbf{I}=(i_{1},i_{2}\cdots)$, $ \mathbf{y}=(y_{1},y_{2},\cdots)$ $ \mathbf{y^{\mathbf{i}}}=\prod_{k\geq 1}y_{k}^{i_{k}}$, and $ m_{\gamma}({\mathbf {I}},x ^{j},z^{n})$ is the number of rooted maps in a surface, with $n$ edges, $j$ faces and $i_{k}$ vertices of valence (degree) $k$. It was shown in \cite{Harer1}, that  $ {\grgen}^s_g (\gamma)$ and $ M_{\gamma}(\mathbf{y},x,z) $ are related through the following formula $$\xi^s_g(\gamma) = s!(-1)^{s}[N^{s}t^{g+s-1}]\Psi M_{\gamma } ,$$   $\Psi$ is the operator defined by $$ \Psi f(\mathbf{y},x,z)=\frac{2}{2}\int_{0}^{1}f(\mathbf{\mathbf(u(t))},x,z)\frac{dz}{z},$$ where $\mathbf{\mathbf u(t)} =(u_{1},u_{2},\cdots)$, $ u_{1}=u_{2}=0 $,  $u_{k}= -i(\sqrt{t})^{k-2}$, $k\geq 3$,  physically means that we respectively discard the tadpole and the self-energy insertions.
 
 \ \  These results show that  $ {\grgen}^s_g (\gamma)$ may be expressed  as coefficients in the generalized Penner free energy, or equivalently  as an  alternating summation for the number of rooted maps; $ m_{\gamma}({\mathbf {I}},x ^{j},z^{n})$. It was conjectured that the number of rooted maps is a polynomial in $1/\gamma$ with integer coefficients \cite{Harer1}.  It is also possible to obtain $ {\grgen}^s_g (\gamma)$  through the action of  the puncture operator $ \frac{1}{s!}\frac{\partial^{s}}{\partial \mu ^{s}}$ ($\frac{1}{s!}\frac{\partial^{s}}{\partial t^{s}}$) on the expression for the free energy in the continuum limit of the generalized Penner model (the Gaussian $\beta$-ensemble), respectively.  The last statement is a consequence of the Penner model because  differentiating  the free energy $n$-times with respect to the continuum variable $\mu$, bring back the punctures to the Riemann surface.
 
 \ \  Now,  identifying our  results for odd-$g$ $ {\grgen}^s_g (\gamma)$ with those in \cite{Harer1}, one reaches the following  equality
\begin{eqnarray}
\label{p20}
 \xi^s_g(\gamma)&=&(-1)^s\frac{(g+s-2)!}{(g+1)(g-1)!}B_{g+1}\nonumber\\&+&(-1)^{s}\frac{(g+s-2)!}{g!}q^{g-1}\left(\sum_{i=1}^{(q-1)/2}(q-2i)B_{g}(\frac{i}{q})\right),\nonumber\\&=&\frac{{(g+s-2)!(-1)^{s+1}} }{ {(g+1)!}}
\left\{ (g+1)B_g q^g+
\sum_{r=0}^{g+1}\binom{ g+1}{r}B_{g+1-r}{B_{r}
}{q^{r}}\right\}.
 \end{eqnarray}
 The last expression in the above equation  corresponds to  the parametrized Euler characteristic derived in \cite{Harer1}.
 If $g=1$, the following formula is deduced
 \begin{eqnarray}
\label{p21}
\sum_{i=1}^{(q-1)/2}(q-2i)B_{1}(\frac{i}{q})=-\left(\frac{1}{12}q^{2}-\frac{1}{4}q+\frac{1}{6}\right).
\end{eqnarray}
On the other hand, and for odd $g$ ($g>1$)
 \begin{eqnarray}
\label{p22}
\sum_{i=1}^{(q-1)/2}(q-2i)B_{g}(\frac{i}{q})=-q^{1-g}\left(B_{g+1}+\frac{1}{g+1}\sum_{r=1}^{g+1}\binom{ g+1}{r}B_{g+1-r}{B_{r}
}{q^{r}}\right),
\end{eqnarray}
from which the following interesting identity \footnote{We very recently proved this identity. The work is in progress. } is obtained
\begin{eqnarray}
\label{identity1}
\sum_{r=1}^{2g}\binom{ 2g}{r}B_{2g-r}B_{r}q^{r}&=&\sum_{r=1}^{g}\binom{ 2g}{2r}B_{2g-2r}B_{2r}q^{2r}\nonumber\\&=&(1-2g)B_{2g}-(2g)q^{2g-2}\sum_{i=1}^{(q-1)/2}(q-2i)B_{2g-1}(\frac{i}{q}).
\end{eqnarray}
Since the second sum given in Eq. (\ref{p19}) has contributions  for $q\geq 3$ only, we should have
\begin{eqnarray}
\label{identity2}
\sum_{r=1}^{2g}\binom{ 2g}{2r}B_{2g-2r}B_{2r}&=&\sum_{r=1}^{g}\binom{ 2g}{2r}B_{2g-2r}B_{2r}2^{2r}\nonumber\\&=&(1-2g)B_{2g}.
\end{eqnarray}
These are well known  formulae for Bernoulli numbers.
The consistency of the formulas given by Eq. (\ref{p21}) and Eq. (\ref{p22})  can be checked 
through the following simple  examples. The first formula for $q=3$ and $q=4$ gives $ B_{1}(1/3)=-1/6$ and $B_{1}(1/4)=-1/4$, respectively.  
Setting $g=3$, $q=3$ and $q=4$, $B_{3}(1/3)= 1/27$ and  $B_{3}(1/4)= 3/64$, respectively. This is  in agreement with the direct evaluation of the Bernoulli polynomials at these rational values.

\ \ Finally, if one recalls the Almkvist-Meurman theorem \cite{Almkvist}, which states that the product  $ q^gB_{g}(i/q)$  for odd $g$ ( $g>1$) and $ 0\leq i\leq q$ is an integer, 
\begin{equation} 
\label{integer}2 q^{g-1}\sum_{i=1}^{(q-1)/2}iB_{g}(\frac{i}{q})-\left(B_{g+1}+\frac{1}{g+1}\sum_{r=1}^{g+1}\binom{ g+1}{r}B_{g+1-r}{B_{r}
 }{q^{r}}\right),
 \end{equation} 
 must be an integer. 
\section {The  generalized Penner model and the Gaussian $\beta$-ensemble}

\ \ We have seen in the last section that the parametrized Euler characteristic  $ \xi^s_g(\gamma)$ for odd $g$ exhibits a contribution from the Euler characteristic of moduli spaces of complex algebraic curves,  thus giving  strong evidence  for  an underlying geometrical meaning. It was suggested in \cite{Harer1} that $ \xi^s_g(\gamma)$ may be  considered as the virtual Euler characteristic of some moduli
spaces, as yet unidentified. Since the generalized Penner model is  a $\gamma$-deformation of the Penner model itself,  one would expect that in the continuum limit (double-scaling), the  free energy that computes the  parametrized Euler characteristic  $ \xi^s_g(\gamma)$   is related to the $c=1$ string theory at radius $R=\gamma$ \cite{Gross}. Also, in the continuum limit, the free energy  of this model  corresponds to the large $N$ asymptotic expansion of the Gaussian $\beta$-ensemble free energy. Before taking the continuum  limit of the generalized Penner model we first review briefly the Gaussian $\beta$-ensemble and then proceed by showing, in detail,  how this model and the generalized Penner model are related to each other. 
 \subsection{ The Gaussian  $\beta$-ensemble}
 
 \ \  The Gaussian  $\beta$-ensemble is defined  by the following partition function \cite{Brini},\cite{Dijkgraaf}
\begin{equation}
 \label{Gaus1}
 Z=\frac{1}{N!(2\pi)^{N}}\int\prod_{i=1}^{N} d\lambda_{i}\vert\Delta(\lambda)\vert^{2\beta}e^{-\frac{\beta}{g_{s}}\sum_{i=1}^{N}\lambda_{i}^{2}},
 \end{equation}
 where $g_{s}$  is the perturbative expansion parameter. This partition function is a deformation of  the Gaussian ensemble partition function by the parameter $\beta$. For finite $N$, the above  matrix integral  can be evaluated  using Mehta's formula \cite{Mehta}
\begin{equation}
\label{Gaus2}
 \int\prod_{i=1}^{N} d\lambda_{i}\vert\Delta(\lambda)\vert^{2\beta}e^{-\frac{1}{2}\sum_{i=1}^{N}\lambda_{i}^{2}}=(2\pi)^{N/2}\prod_{k=1}^{N}\frac{\Gamma(1+\beta k)}{\Gamma(1+\beta )}. 
\end{equation}
Setting $\beta=1$, $ Z\thicksim
\prod_{k=1}^{N-1} k!\thicksim 1/vol(U(N))$. Here,  $vol(U(N))$ is  the volume of the unitary gauge group  for the partition function of the Gaussian ensemble,
$$  Z=\frac{1}{vol(U(N))}\int dM e^{-\frac{1}{g_{s}} Tr M^{2}},$$  where the integration is over $N\times N$ Hermtiam matrix $M$.  The expression $ \prod_{k=1}^{N-1} k!$ is the Barnes gamma function $G_{2}(z)$  defined by $G_{2}(N+1)=\prod_{k=1}^{N-1} (N-k)!$, known in the large $N$ expansion to reproduce all the genera contributions of the $B$-model on the conifold \cite{Ooguri}. This is also the Penner model in the continuum limit. The special values $ \beta=2, 1/2$ compute the volume of the gauge groups $ Sp(N)$ and  $SO(N)$ respectively. In the large $N$ limit, these volumes give rise to the $ Sp(N)/ SO(N)$ Penner models in the continuum limit \cite{Chair2}.

\ \   For $\beta=2$, the Gaussian partition function can be written as $$ Z\thicksim
\frac{1}{N!}\prod_{k=1}^{N} \frac{(2k)!}{2}= \prod_{k=1}^{N}(2k-1)!\thicksim \frac{1}{vol(Sp(2N))},$$ while for $\beta=1/2$ and using the Legendre duplication formula, the partition function reads

$$Z\thicksim \frac{1}{N!}\prod_{k=1}^{N}\frac{\Gamma(1+k/2)}{1/2 \sqrt{\pi}}\thicksim(N-2)!(N-4)!\dots 6!.4!.2!\thicksim  \frac{1}{vol(Sp(N-1))},$$ where $N$ is assumed to be even. It was shown \cite{Chair2} that $vol(Sp(2N-1)) $ and  $vol(SO(2N)) $ are equivalent, thus, $Z\thicksim  \frac{1}{vol(SO(N))}$ for $\beta=1/2$.

\ \  This very close relationship between the Gaussian $\beta$-ensemble and the generalized Penner model is  expected though. The  partition function for the  Gaussian $\beta$-ensemble given by Eq. (\ref{Gaus1}) plays the role of the volume for certain gauge group of the generalized Penner model partition function given by Eq. (\ref{p0}). Here, we used the fact that the matrix models are gauge theories like the Chern-Simons gauge theories, and  nonperturbative terms in such models are captured by the volume of the corresponding gauge groups. One should also point out that the nonpertubative terms which are reproduced in the double scaling limit of the Chern-Simons gauge theory may be extended  to matrix models as well. 
\subsection{The double scaling limit of the generalized Penner Model}

\ \  We  will show that in the continuum limit,  this model reproduces the generating function for the parametrized  Euler characteristic without punctures,  and  has  the same critical points as the Penner model \cite{Chair3}, \cite{Vafa1}. To that end, let us  write  the  free energy for the generalized Penner model as
\begin{eqnarray}
\label{cont1}
F_{q}(N,t)&=\frac{1}{q}\sum_{g,s}\frac{(-1)^{s}}{s!}\xi^{s}_g(\gamma) N^{s}t^{g+s-1},
\end{eqnarray}
where the natural scaling $ t\rightarrow t/N $ is used. The parametrized Euler characteristic $ \xi^{s}_g(\gamma)$ splits into two sectors of even and odd $g$. Therefore, $F_{q}(N,t)=F^{e}_{q}(N,t)+F^{o}_{q}(N,t) $, where $F^{e}_{q}(N,t)$ and $F^{o}_{q}(N,t)$ refer to the even and odd $g$ contributions  to the free energy, respectively.  For even $g$, the contribution to the free energy  reads
\begin{eqnarray}
\label{cont2}
F^{e}_{q}(N,t)&=&\sum_{g,s}\frac{(2g+s-2)!}{s!2}\left(q^{2g-1}-1\right)\frac{B_{2g}}{(2g)!}N^{1-2g}t^{2g-1+s}\nonumber\\&=&\frac{N}{2}\sum_{s=2}\frac{t^{s-1}}{s(s-1)}\left(\frac{1}{q}-1\right)\nonumber\\&+&\frac{1}{2}\sum_{g\geq 1}\sum_{s\geq 0}\frac{(2g+s-2)!}{s!}\left(q^{2g-1}-1\right)\frac{B_{2g}}{(2g)!}N^{1-2g}t^{2g-1+s}.
\end{eqnarray}

\ \ The sum over punctures is carried out and gives 
 \begin{eqnarray}
\label{cont3}
F^{e}_{q}(N,t)&=&\frac{N}{2}\Bigr{(}1+(\frac{1-t}{t})\log(1-t)\Bigl{)}\left(\frac{1}{q}-1\right)\nonumber\\&+&\frac{1}{2}\sum_{g\geq 1}\Bigl(\frac{N(1-t)}{t}\Bigr)^{1-2g}\left(q^{2g-1}-1\right)\frac{B_{2g}}{(2g)(2g-1)}.
\end{eqnarray}
To obtain the continuum limit for the free energy $F_{q}(N,t)$, one sets $ \mu=N(1-t)$, $N\rightarrow \infty$ and $t \rightarrow 1$, such that $ \mu$ is kept fixed (double scaling limit) and gets
\begin{eqnarray}
\label{cont4}
F^{e}_{q}(\mu)&=&\frac{\mu}{2}\log\mu\left(\frac{1}{q}-1\right)\nonumber\\&+&\frac{1}{2}\sum_{g\geq 1}{\mu}^{1-2g}\left(q^{2g-1}-1\right)\frac{B_{2g}}{(2g)(2g-1)}.
\end{eqnarray}
This  is the generalization of  the orthogonal  free energy Penner model in the continuum limit ($q=2$) \cite{Chair1}, \cite{Chekhov}, with the same critical points as the Penner model \cite{Chair3}, \cite{Vafa1}.

\ \ For odd $g$,  the free energy contribution is 
\begin{eqnarray}
\label{cont5}
F^{o}_{q}(N,t)&=&\frac{1}{q}
\sum_{g,s}\frac{(2g+s-3)!}{s!(2g)!}(2g-1)B_{2g}N^{2-2g}t^{2g-2+s}\nonumber\\&+&\sum_{g,s}\frac{(2g+s-3)!}{s!(2g-1)!}q^{2g-3}\left(\sum_{i=1}^{(q-1)/2}(q-2i)B_{2g-1}(\frac{i}{q})\right)N^{2-2g}t^{2g-2+s},
 \end{eqnarray}
 where the first term is the free energy of the  Penner model discussed in detail elsewhere  \cite{Chair3}, while the second term is the  free energy contribution  for $q\geq 3$. Summing the latter term over the punctures gives
\begin{eqnarray}
\label{cont6}
&&\sum_{g,s}\frac{(2g+s-3)!}{s!(2g-1)!}q^{2g-3}\left(\sum_{i=1}^{(q-1)/2}(q-2i)B_{2g-1}(\frac{i}{q})\right)N^{2-2g}t^{2g-2+s}\nonumber\\&&=\log(1-t)\left( \frac{1}{12}q-\frac{1}{4}+\frac{1}{6q}\right)+\nonumber\\&&\sum_{g\geq 2}\Bigl(\frac{N(1-t)}{t}\Bigr)^{2-2g}\frac{q^{2g-3}}{(2g-1)(2g-2)}\left(\sum_{i=1}^{(q-1)/2}(q-2i)B_{2g-1}(\frac{i}{q})\right).\nonumber\\
\end{eqnarray}

\ \ Therefore, the contribution to the free energy   reads
\begin{eqnarray}
\label{cont7}
F^{o}_{q}(\mu)&=&\frac{1}{2q}\mu^2 \log\mu-\frac{1}{12q}\log\mu+\frac{1}{q}\sum_{g\geq2}\frac{1}{(2g-2)}\frac{B_{2g}}{2g}\mu^{2-2g}\nonumber\\&+&\frac{1}{q}\left( \frac{1}{12}q^2-\frac{q}{4}+\frac{1}{6}\right)\log\mu\nonumber\\&+&\frac{1}{q}\sum_{g\geq 2}\frac{1}{(2g-1)(2g-2)}\left(\sum_{i=1}^{(q-1)/2}(q-2i)B_{2g-1}(\frac{i}{q})\right)(\mu/q)^{2-2g}.
\end{eqnarray}
 If we set $q=2$, we recover our previous results \cite{Chair1} on the orthogonal Penner model in which the orientable contribution part gives half the Penner free energy.
 
 \ \  Our expression for the  free energy $F_{q}(\mu)$ has two sectors; the even and the odd powers in $\mu$. This is reminiscent of  the  large $N$ asymptotic expansion of the Gaussian $\beta$-ensemble free energy \cite  {Walcher,Brini, Krefl}.  Discarding the regular terms in $\mu$ and  adding  Eqs. (\ref{cont5}) to (\ref{cont7}), the total free energy  is
 \begin{eqnarray}
\label{Gbeta}
\mathcal{F_{\gamma}(\mu)}&=&\frac{\gamma\mu^{2}}{2 }\log\mu +\frac{\gamma -1}{2}\mu\log\mu +\frac{1-3\gamma+ \gamma^{2}}{12\gamma}\log\mu\nonumber\\&+&\frac{1-\gamma}{24\gamma\mu}+ \frac{1-5\gamma^{2}+ \gamma^{4}}{720\gamma^{3}\mu^{2}}+\frac{1-\gamma^{3}}{720\gamma^{3}\mu^{3}}+\cdots\nonumber\\&+& \big(\frac{\gamma}{10080}-\frac{1}{3040\gamma}-\frac{1}{4370\gamma^{3}}-\frac{1}{3040\gamma^{5}}+\frac{1}{10080\gamma^{7}}\big)\frac{1}{\mu^{6}}\nonumber\\&+&\frac{1}{2}\sum_{g\geq 3}\left(\frac{1}{\gamma^{2g-1}}-1\right)\frac{B_{2g}}{(2g)(2g-1)}{\mu}^{1-2g}\nonumber\\&-&\gamma\sum_{g\geq5}\frac{1}{(2g-2)(2g-1)}\frac{B_{2g}}{2g}\mu^{2-2g}\nonumber\\&-&\gamma \sum_{g\geq 5}\frac{1}{(2g-2)(2g-1)2g} \left(\sum_{r=1}^{2g}\binom{ 2g}{r}B_{2g-r}{B_{r}
}{(\frac{1}{\gamma})^{r}}\right)\mu^{2-2g}.
\end{eqnarray}

\ \  Let us now compare our results with those  obtained in \cite{Brini} for the free energy of the Gaussian $\beta$-ensemble in the large $N$ limit. In \cite{Brini}, the   partition function  for the Gaussian  $\beta$-ensemble for finite $N$ is expressed in terms of the double Gamma Barnes function $ \Gamma_{2}(x;a,b)$. Using the asymptotic expansion for $ \log\Gamma_{2}(x;a,b)$ up to some additive terms \cite{Spreafico}, the expression for the free energy \cite{Brini} reads
\begin{eqnarray}
\label{doublegam}
F=-\log\Gamma_{2}(t;-g_{s},g_{s}/\beta)&=&\frac{\beta t^{2}}{2 }(\log t -\frac{3}{2})g_{s}^{-2}+\frac{\beta -1}{2}t(\log (\beta t)-1)g_{s}^{-1} \nonumber\\&+&\frac{1-3\beta + \beta^{2}}{12\beta}\log( \beta t)+\frac{1-\beta}{24\beta t}g_{s}\nonumber\\&+& \frac{1-5\beta^{2}+\beta^{4}}{720\beta^{3}t^{2}}g_{s}^{2} +\frac{1-\beta^{3}}{720\beta^{3}t^{3}}g_{s}^{3}+ \cdots ,
\end{eqnarray}
where $t$, $g_{s}$ are the 't Hooft coupling and  the string coupling constant respectively. 
This expression  shows clearly that if we let $ \gamma=\beta$ and  $\mu=\frac{t}{g_{s}}$, up to regular terms, the coefficients of $ \mathcal{F_{\gamma}(\mu)}$ are exactly those of the Gaussian $\beta$-ensemble free energy for large $N$ \cite{Brini}. This also shows
 that one may consider the partition function for the Gaussian $\beta$-ensemble  as the volume of certain gauge group which we do not know in general. For $ \beta=2, 1/2$, the gauge groups are $ Sp(N)$ and  $ SO(N)$,  respectively.  Note that the coefficients of  even and odd powers of $\mu$ are separated in our formula  from each other, unlike those of the  Gaussian $\beta$-ensemble. 
\section{Duality symmetry of the generalized Penner model and the Gaussian $\beta$-ensemble}

\ \ It was shown in \cite{Mulase}, that the free energy of the generalized Penner model bears the duality $ F_{\gamma}(N,t)=F_{\frac{1}{\gamma}}(-\gamma N,t)$. This duality  first appeared in connection with the equivalence of $Sp(2N)$ and $SO(-2N)$ gauge theories\cite{Mkrtchyan}.  It is  interesting to note that such symmetry still holds in the continuum limit  for all terms without logarithmic singularities albeit $N$ is replaced by the parameter $\mu$, that is, $\mathcal{F_{\gamma}(\mu)}=\mathcal {F}_{1/\gamma}(-\gamma \mu)$.  Although the logarithmic singularities break this duality,   their coefficients do enjoy  it.

\ \ This duality  manifests itself clearly in the odd sector, while  for the even sector, the duality may be shown to hold by realizing that  the last term but one in Eq. (\ref{Gbeta}) is dual to the last term with $r=2g$. The other terms that are dual to each other are those terms whose coefficients are  $\binom{ 2g}{2r}B_{2g-2r}{B_{2r}}=\binom{ 2g}{2g-2r}B_{2r}{B_{2g-2r}}$, $ B_{r}=0$ for odd  $r>1$. There is also a self-dual term whose  coefficient is $ \binom{ 2g}{g}B_{g}^2$. As a consequence, this duality  is also  a duality symmetry of the Gaussian $\beta$-ensemble free energy for large $N$.

\ \ The duality transformation of the Gaussian $\beta$-ensemble should be $\beta \rightarrow {\frac{1}{\beta}}$ and $ t \rightarrow -\beta t$. This fact can be easily checked  using  the non-logarithmic terms in Eq. (\ref{doublegam}). In order to test this duality term by term  in  the free energy, one needs to know the explicit  expression  for the terms in the  free energy  $ F=-\log\Gamma_{2}(t;-g_{s},g_{s}/\beta)$ that are neither constants nor logarithmic.  Using the asymptotic expansion of the double Gamma Barnes function  \cite{Spreafico}, these terms are  
\begin{eqnarray}
\label{duality1}
\sum_{n=1}^{\infty}(n-1)!e_{n}(a,b)t^{-n},
\end{eqnarray}

 where  the coefficients $e_{n}(a,b) $  for  $ a=-g_{s}$ and $ b= g_{s}/\beta $ are;
 \begin{eqnarray}
\label{duality}
  e_{n}(-g_{s},g_{s}/\beta)&=& \frac{B_{n+1}}{2(n+1)!}\Big((-1)^{n}+\frac{1}{\beta^{n}}\Big)g_{s}^{n}-\frac{B_{n+2}}{(n+2)!}\Big((-1)^{n}+\frac{1}{\beta^{n+2}} \Big)\beta g_{s}^{n}\nonumber\\&-&
  \sum_{j=0}^{n-2}(-1)^{n-j}\frac{B_{n-j}B_{j+2}}{(n-j)!(j+2)!} (1/\beta)^{j+1}g_{s}^{n}, \text{for } n \geq 2.  
 \end{eqnarray}
 At this point, it is not difficult to see that the first two terms in $(n-1)!e_{n}(-g_{s},g_{s}/\beta)t^{-n} $ are invariant under the  transformation $\beta \rightarrow {\frac{1}{\beta}}$ and $ t \rightarrow -\beta t$.  Under this  transformation, the last term may be written as $$ -\sum_{j=0}^{n-2}(-1)^{j}\frac{B_{n-j}B_{j+2}}{(n-j)!(j+2)!} (\beta)^{j+1}g_{s}^{n} (\beta t)^{-n}.$$  A simple exercise shows that this is exactly the last term $$ -\sum_{j=0}^{n-2}(-1)^{n-j}\frac{B_{n-j}B_{j+2}}{(n-j)!(j+2)!} (1/\beta)^{j+1}g_{s}^{n}t^{-n}.$$ As in the previous derivation of the duality of the generalized Penner model,  there is a self-dual term for  $j=\frac{n-2}{2} $, for $n$ even. If  $n$  is odd, the above sum is identically zero since $B_{m}=0$, for  odd $m\geq 3$. Therefore, we conclude that the duality symmetry of the  Gaussian $\beta$-ensemble is a consequence of the duality symmetry of the generalized Penner model.
 
 \ \  Note that  the above  results would have been obtained if we were to use the Schwinger integral \cite{Walcher}. In this representation, the free energy of the Gaussian  $\beta$-ensemble can be written as
 $$ F^{G}\thicksim \int\frac{ds}{s} \frac{e^{-ts}}{(e^{\epsilon_{1}s}-1)(e^{\epsilon_{2}s}-1)}\thicksim \Phi^{0}(\beta)\log(t)+\sum_{n\geq 1} \frac{g_{s}^{n}}{t^{n}}\Phi^{n}(\beta),$$
 where $\epsilon_{1}=g_{s}\sqrt{\beta}$, $\epsilon_{2}=-\frac{g_{s}}{\sqrt{\beta}}$. $\Phi^{0}(\beta)$ and   $\Phi^{n}(\beta)$ are the  $\beta$ dependent coefficients that appeared in the double Gamma Barnes function. The above expression for the free energy is obviously symmetric under the exchange $\epsilon_{1}\longleftrightarrow \epsilon_{2}$. However, the expression is generally not invariant under the exchange $(\epsilon_{1},\epsilon_{2})\rightarrow (-\epsilon_{2}, -\epsilon_{2}) $ except when $ \epsilon_{1}=-\epsilon_{2}$. Making a shift $t\rightarrow t+ \frac{\epsilon_{1}+\epsilon_{2}}{2}$, this symmetry is restored.  The asymptotic expansion in this case is
 $$ F^{G}(t;g_{s},\beta)=\int\frac{ds}{s} \frac{e^{-ts}}{(e^{\epsilon_{1}s}-e^{-\epsilon_{1}s})(e^{\epsilon_{2}s}-e^{-\epsilon_{2}s})}\thicksim \Psi^{0}(\beta)\log(t)+\sum_{n\geq 1} \frac{g_{s}^{n}}{t^{n}}\Psi^{n}(\beta),$$
 where 
 \begin{eqnarray}
\label{c=1}
 \Psi^{0}(\beta)&=&-\frac{1}{24}(\beta+\beta^{-1}),
\end{eqnarray}
and for $n\geq 1$
\begin{eqnarray}
\label{c=2}
\Psi^{n}(\beta)&=&(n-1)!\sum_{j=0}^{n+2}(-1)^{j}\frac{B_{n+2-j}B_{j}}{(n+2-j)!(j)!}(2^{1-j}-1) (2^{1-n-2+j}-1)(\beta)^{j-n/2-1}.
 \end{eqnarray}
 
 \ \  The right-hand side is zero for $n$ odd, and since $B_{j}\neq 0$ for  $ j$ even, the free energy without logarithmic singularities reads
\begin{eqnarray}
\label{c=3} 
 F^{G}(t;\beta,g_{s}) &\thicksim & \sum_{m=1}^{\infty} (2m-1)!\sum_{j=0}^{m+1}\frac{B_{2m+2-2j}B_{2j}}{(2m+2-2j)!(2j)!}\nonumber\\&\times& (2^{1-2j}-1) (2^{1-2m-2+2j}-1)(\beta)^{2j-m-1}\frac{g_{s}^{2m}}{t^{2m}}.
\end{eqnarray}
The above equation shows  that $  F^{G}(t;\beta ,g_{s})= F^{G}(-t\beta;1/\beta ,g_{s})$ under the duality transformation  ($\beta \rightarrow {\frac{1}{\beta}}$ and $ t \rightarrow -\beta t$) since  the sum over $j$ remains invariant.  That is, the duality transformation is preserved when  shifting   the  coupling;  $t\rightarrow t+ \frac{\epsilon_{1}+\epsilon_{2}}{2}$. Under this shift, the powers of $t$ in the expansion for the free energy  are even, so the duality in this case is $\beta \rightarrow {\frac{1}{\beta}}$ and $ t \rightarrow \beta t$. This is exactly the remarkable duality symmetry obtained for the $c=1$ string at radius $\beta$ \cite{Gross}.
\section{Conclusion}

\ \ In this work alternative  new formulas  for the partition function as well as the free energy that computes the  Parametrized Euler Characteristic of the generalized Penner model are given. These formulas contain both the Penner and the orthogonal Penner models as special cases. Furthermore, for odd $g$ and  for all the parameters $\gamma$, the parametrized Euler characteristic  exhibits a contribution from the orbifold Euler characteristic of Riemann surfaces of genus $g$ with $s$ punctures.

\ \  Our explicit formula for the free energy of the generalized Penner model gives exactly the same coefficients as  the Gaussian $\beta$-ensemble free energy in the large $N$ limit. Also our formula shows clearly that the duality enjoyed by the generalized Penner model is preserved in the continuum limit,  albeit $N$ is replaced by the coupling $\mu$, $\mathcal{F_{\gamma}(\mu)}=\mathcal {F}_{1/\gamma}(-\gamma \mu)$. This duality in turn induces the same duality for the Gaussian  $\beta$-ensemble; $\beta \rightarrow {\frac{1}{\beta}}$ and $ t \rightarrow -\beta t$.  This duality symmetry survives  the shift of the coupling constant $t$ required by the refined topological string. The duality in this case coincides  with that of  the $c=1$ string at radius $\beta$. In terms of the equivariant parameters $\epsilon_{1}= \sqrt{\beta}g_{s}$, $\epsilon_{2}= -\frac{g_{s}}{\sqrt{\beta}}$,  $\beta=-\frac{\epsilon_{1}}{\epsilon_{2}}$, this duality may be written as $\epsilon_{1}\longleftrightarrow \epsilon_{2}$, and  $ t \rightarrow \frac{ \epsilon_{1}}{\epsilon_{2}}t$.

\ \  We have  recently shown that in the continuum limit  both the $SO$ Chern-Simons gauge theory \cite{Vafa2} and the $SO$ Penner  model are equivalent  \cite {Chair2}. Therefore, we may ask if there is  a Chern-simons gauge theory whose free energy in the continuum limit (the logarithm of the volume of the gauge group in the large $N$ limit), is  given by  Eq. (\ref{Gbeta}). This may correspond to a topological string on the quotient of the resolved conifold by the discrete group $Z_{\gamma}$. 
\appendix
\section{Derivation of the proposed identity.}
\label {identity}

\ \ We present  an  explicit derivation of the proposed identity given in Eq. (\ref{p8}). The  multiple product of the left-hand-side of  Eq. (\ref{p8}) may be expanded to give 
$$\frac{{\prod_{l=1}^K\prod_{j=1}^{ql}(1-jt)}
}{ {\prod_{j=1}^K(1-tqj)}}= \prod_{j=1}^{q}(1-jt)\prod_{j=1}^{2q}(1-jt)\cdots \prod_{j=1}^{qK-q}(1-jt)\prod_{j\nmid{q}}^{qK-1}(1-jt),$$ where $$ \prod_{j\nmid{q}}^{qK-1}(1-jt)=\frac{\prod_{j=1}^{qK}(1-jt)}{\prod_{j=1}^K(1-tqj)}$$ here, $j\nmid{q}$ means that $j$ is not multiple of 
$q$, and $N=qK$.  The above product is reminiscent of certain products  connected with the Euler gas \cite{Chair4}.  Explicitly, this product may be written as  $$\prod_{j\nmid{q}}^{qK-1}(1-jt)= \prod_{j=1}^{q-1}(1-jt)(1-(q+1)t)\cdots (1-(N-(q-1))t)\cdots (1-(N-1)t), $$  by canceling all the terms in the product for which $j$  is a  multiple of $q$. Combining the $ q-1$  products  $(1-(N-(q-1))t)\cdots (1-(N-1)t) $ with $\prod_{j=1}^{qK-q}(1-jt) $ gives $\prod_{j=1}^{N-1}(1-jt) $. In the same way a term like $\prod_{j=1}^{N-q-1}(1-jt) $ may be obtained by combining  the $ q-1$  products  $(1-(N-(2q-1))t)\cdots (1-(N-(q+1)t) $ with $\prod_{j=1}^{qK-2q}(1-jt) $.  Continuing this process, we obtain the following
\begin{eqnarray}
\label{ID}
\frac{{\prod_{l=1}^K\prod_{j=1}^{ql}(1-jt)}
}{ {\prod_{j=1}^K(1-tqj)}}&=&\prod_{j=1}^{N-1}(1-jt) \prod_{j=1}^{N-q-1}(1-jt)\cdots\prod_{j=1}^{3q-1}(1-jt)\prod_{j=1}^{2q-1}(1-jt)\prod_{j=1}^{q-1}(1-jt)\nonumber\\&=&\prod_{j=0}^{N/q-1}\prod_{p=1}^{N-(qj+1)}(1-pt)
\end{eqnarray}
The last line in Eq. (\ref{p8}) is obtained by separating the product $ \prod_{p=1}^{N-(qj+1)}(1-pt)$ into a product for which $p$ is congruent to $q$ and a product for which $p$ is not congruent to  $q$. That is, $$\prod_{p=1}^{N-(qj+1)}(1-pt)= \prod_{p=1}^{N/q-j}(1-(qp-(q-1))t)\prod_{p=1}^{N/q-j}(1-(qp-(q-2))t)\cdots \prod_{p=1}^{N/q-j-1}(1-(qp)t).$$
The product over $j$ may be carried out by realizing that the term with $p=1$ in all products with the coefficient of $t$ being non-congruent to $q$  appears $ N/q$ times, $p=2$ appears $ N/q-1$ times and $p=i$ appears $ N/q-i+1$ times. The product over $j$ for the last product in which the coefficient of $t$ being congruent to $q$, shows that the term with $p=i$ appears $ N/q-i$ times, q.e.d.
\section{The parametrized Euler characteristic for $g$ even.}
\label {PEc}

\ \ In this appendix we derive the expression for the parametrized Euler characteristic for even $g$ as given in Eq. (18). The first line given in Eq. (\ref{p11}) does  not contribute to the parametrized Euler characteristic $\xi^s_g(\gamma)$, while the contribution of  Eq. (\ref{p14}) is  
\begin{eqnarray}
\label{p15}
\xi^s_g(\gamma)&=&(-1)^{s+1}\frac{(g+s-2)!}{g!}\left(1-q^{g-1}\right)B_{g}\nonumber\\&+&(-1)^{s+1}\frac{(g+s-2)!}{g!}q^{g-1}\left( B_{g}(\frac{2}{q})+2B_{g}(\frac{3}{q})+\cdots+(q-2)B_{g}(\frac{1}{q})\right), 
\end{eqnarray}
Using the symmetry  $B_{g}(1-x)=B_{g}(x)$,  we may write $$ B_{g}(\frac{2}{q})+2B_{g}(\frac{3}{q})+\cdots+(q-2)B_{g}(\frac{1}{q})=\frac{q-2}{2}\left(2B_{g}(\frac{1}{q})+2B_{g}(\frac{2}{q})+\cdots+B_{g}(\frac{q/2}{q})\right), $$
The sum  on the right-hand side can  be written in a closed form by evaluating  the multiplication formula for the Bernoulli polynomials $$ B_{q}(kx)=k^{q-1}\sum_{k=0}^{q-1}\binom{n}{k}B_{q}(x+j/k), $$  at $x=0$. Then, a simple computation shows $$2B_{g}(\frac{1}{q})+2B_{g}(\frac{2}{q})+\cdots+B_{g}(\frac{q/2}{q})=\left(\frac{1}{q^{g-1}}-1\right)B_{g}. $$ 
Therefore, if $g$ is even, the  parametrized Euler characteristic $\xi^s_g(\gamma)$ becomes
\begin{eqnarray}
\label{pPEE}
 \xi^s_g(\gamma)&=&(-1)^{s+1}\frac{(g+s-2)!}{g!}\left(1-q^{g-1}\right)B_{g}\nonumber\\&+&(-1)^{s+1}\frac{(g+s-2)!}{g!}q^{g-1}\left( \frac{q-2}{2} \right)\left(\frac{1}{q^{g-1}}-1\right)B_{g}\nonumber\\&=&(-1)^{s}\frac{(g+s-2)!}{g!2}\left(q^g-q\right)B_{g}
 \end{eqnarray}
 \vspace{7mm}
 
{\bf Acknowledgments:}
I would like to thank G. Bonelli, S. Cecotti and K.S. Narain for discussions and reading the manuscript, J. Walcher for correspondence. Also, I would like to thank Martin O'Loughlin for useful comments  and   the Abdus Salam Centre for Theoretical Physics, Trieste for the supports they give me.  
 
\end{document}